\documentclass[12pt]{article}
\usepackage[dvips]{graphicx}
\usepackage{amsbsy}
\usepackage{amssymb}

\newcommand{\beq}{\begin{equation}}
\newcommand{\eeq}{\end{equation}}
\newcommand{\bdm}{\begin{displaymath}}
\newcommand{\edm}{\end{displaymath}}
\newcommand{\beqr}{\begin{eqnarray}}
\newcommand{\eeqr}{\end{eqnarray}}
\newcommand{\beqrn}{\begin{eqnarray*}}
\newcommand{\eeqrn}{\end{eqnarray*}}

\def\l{\lambda}
\def\bchi{\boldsymbol{\chi}}

\def\l{\lambda}

\begin{document}

\title{Generating functions and multiplicity formulas: the case of rank two simple Lie algebras}

\author{Jos\'e Fern\'andez N\'u\~{n}ez$^{\dagger}$, Wifredo Garc\'{\i}a Fuertes$^{
\ddagger}$\\
\small Departamento de F\'\i sica, 
Facultad de Ciencias, Universidad de Oviedo, 33007-Oviedo, Spain\\
\small {\it $^\dagger$nonius@uniovi.es; $^\ddagger$wifredo@uniovi.es}\\
\and
 Askold M. Perelomov\\
\small Institute for Theoretical and Experimental Physics, Moscow, Russia.\\
\small {\it  aperelomo@gmail.com}}

\date{ }

\maketitle

\begin{abstract}
\noindent
A procedure is described that makes use of the generating function of characters to obtain a new generating function $H$ giving the multiplicities of each weight in all the representations of a simple Lie algebra. The way to extract from $H$ explicit multiplicity formulas for particular weights is explained and the results corresponding to rank two simple Lie algebras shown.
\end{abstract}
\bigskip


\medskip

\vfill\eject 
\section{Introduction}
Each irreducible representation of a simple Lie algebra is defined by a set of weights giving the eigenvalues of the mutually commuting generators of the Cartan subalgebra on the states forming the representation. These states arise from iterated application on the highest weight state of the lowering operators $E_{-\alpha}$ corresponding to the positive roots $\alpha$ of the algebra, but there are, in general, more than one state per weight, i.e. the weights enter in the representation with some multiplicity. Weight multiplicities are basic information to built the representations of the Lie algebra and, from the point of view of physics, they give the degeneracies of the quantum stationary states of the systems whose dynamics has the Lie group as an underlying symmetry. For these reasons, the computation and understanding of weight multiplicities in the representations of simple Lie algebras has been a subject of much research along the years \cite{wi37}--\cite{mopa82}. As it often happens whith many issues having to do with Lie algebra representations, one of the most efficient tools available to tackle the question of multiplicities is the theory of characters. In a recent paper~\cite{nfp14}, we have presented a general method for computing the generating function of the characters of simple Lie algebras by means on the theory of the quantum trigonometric Calogero-Sutherland system \cite{ca71}--\cite{op76} (see also \cite{ps79,ow07} for other approaches to that problem). One advantage of the method described in \cite{nfp14} is its simplicity: it requires some acquaintance with the Calogero-Sutherland theory, especially its treatment by means of Weyl-invariant variables but, apart from that, the computations involved are quite elementary.

Based on the results of \cite{nfp14},  we have devised in a recent paper \cite{nfp15}  a method to obtain the generating function for weight multiplicities,  and applied it  specifically to the case of the Lie algebra $C_2$. In the present paper our aim is to obtain formulas for the weight multiplicities for all simple Lie algebras of rank two,  i.e. for the cases of $A_2$, $B_2$, $C_2$ and $G_2$. For each algebra,
we will take advantage of the generating function for characters to obtain a generating function $H$ which collects in a single object all the information about multiplicities for all weights and all the representations of the algebra, and we will explain how to use $H$ to obtain explicit formulas giving the multiplicities of any particular weight in all the representations. In themselves the computations are rather straightforward, although the algebra can be somewhat convoluted due to the need to manipulate some long polynomials or rational expressions. One is thus forced to use some computer program for symbolic calculus, but the required level of performance on computer methods is by not means demanding and the approach should be accessible to everyone with a little familiarity with standard programs like Mathematica or Maple.

 Due to the low rank of the algebras we are dealing with, these are of course the most natural cases to be treated in order to provide some illustrations of the approach. Nevertheless, the rank two algebras are interesting per se by virtue of their role in several important physical situations. This is especially true for $A_2$, which provides the classification of baryons in the Eightfold Way, is the gauge group of QCD and appears in the compactifications of superstring theory as the required holonomy group to give $N=1$ supersymmetry in four dimensions. In the case of the isomorphic algebras $B_2$ and $C_2$, there are physical applications in nuclear physics \cite{rw13}, the theory of high temperature superconductors \cite{zh97} or string and M-theory \cite{gi02,bbs07}. And with regard to the exceptional algebra $G_2$, let us mention its use in atomic physics to explain some spectroscopical anomalies \cite{ra52}, in gauge field theory as a very convenient model for lattice simulations \cite{mw12}, or in the compactifications of M-theory to four dimensions \cite{aw03}. In view of the above applications, the formulas to be given in the Section 3 of the paper should be considered not only as mere examples of the approach, but also as results with the potential to be useful in some interesting physical contexts.
\section{Description of the method}
We will explain how the calculations go focusing only in the case of a rank two simple Lie algebra, given that these are the examples to be studied later in detail, but the generalization to higher rank algebras is entirely obvious   \cite{nfp15}. Thus, let ${\cal A}$ be such an algebra with fundamental weights $\lambda_1$ and $\lambda_2$ and let us consider the irreducible representation  $R_{p \lambda_1+q \lambda_2}$ of ${\cal A}$ with highest weight $\lambda=p \lambda_1+q \lambda_2$. The character of this representation is defined as 
\beq
\bchi_{p,q}=\sum_{m,n} \mu_{p,q}(m,n)\, e(m,n)\,, \label{eq:car}
\eeq
where  the sum extends to all weights $w=m \lambda_1+n \lambda_2$ entering in the representation, $\mu_{p,q}(m,n)$ is the  multiplicity of the weight $w$ and $e(m,n)$ is the exponential $e(m,n)=x_1^{m} x_2^{n}$, with $x_1$ and $x_2$ being complex phases, $x_l=e^{i \varphi_l}$, and where $\varphi_1,\varphi_2$ are angular coordinates on the maximal torus. It follows from this definition that the multiplicity $\mu_{p,q}(m,n)$ of the weight $w$ in the representation $R_{p \lambda_1+q \lambda_2}$ can be computed by means of the integral
\beq
\mu_{p,q}(m,n)=\frac{1}{(2\pi i)^2}\oint d x_1\oint d x_2 \frac{\bchi_{p,q}}
{x_1^{1+m} x_2^{1+n}}\, , \label{eq:nmp}
\eeq
where the integration contours are the unit circles on the complex planes parametrized by the coordinates $x_1$ and $x_2$. Let us now introduce the generating function for the characters of the algebra,
\beq
G(t_1,t_2; z_1,z_2)=\sum_{{p,q=0}}^\infty t_1^{p}\, t_2^{q}\, \bchi_{p,q}(z_1,z_2)\ , \label{eq:gcar}
\eeq
which appears here expressed in terms of the characters of the fundamental representations \mbox{$z_1=\bchi_{1,0}$}, $z_2=\bchi_{0,1}$, a suitable set of variables to compute it in closed form. Let us pay attention to the integral 
\beq
H(t_1,t_2;y_1,y_2)=\frac{1}{(2\pi i)^2}\oint d x_1\oint d x_2 
\frac{G(t_1,t_2;z_1,z_2)}{(x_1-y_1)(x_2-y_2)}\ . \label{eq:h}
\eeq
By trading the denominator by a product of geometric series of ratios ${y_1}/{x_1},\ {y_2}/{x_2}$ and comparing with (\ref{eq:nmp}) and (\ref{eq:gcar}), we see that the coefficient of $t_1^{m}\,t_2^{n}\,y_1^{p}\,y_2^{q}$ in the resulting expansion is nothing else but the multiplicity of the weight $w=m \lambda_1+n \lambda_2$ in the representation of $R_{p\lambda_1+q\lambda_2}$. Of course, the multiplicity is preserved in a Weyl reflection, and all weights entering in $R_{p\lambda_1+q\lambda_2}$ are images by the Weyl group of a dominant weight with $m, n \geq 0$. It follows that $H(t_1,t_2;y_1,y_2)$ is a quite comprehensive generating function which collects in itself the multiplicities of all weights entering in all the representations of the algebra ${\cal A}$. In particular, if we consider a partial expansion in the $y_1$, $y_2$ variables only,
\beq
\label{HA}
H(t_1,t_2;y_1,y_2)=\sum_{m,n=0}^\infty A_{m,n}(t_1,t_2) \,y_1^m \,y_2^n\,,
\eeq
we obtain a series whose coefficients are themselves the generating functions $A_{m,n}(t_1,t_2)$ which give the multiplicity of a particular dominant weight in all the representations of the algebra:
\beq
\label{Amu}
A_{m,n}(t_1,t_2)=\sum_{p,q=0}^\infty\mu_{p,q}(m,n) t_1^p\, t_2^q \ .
\eeq
Conversely, the coefficient of $t_1^p\, t_2^q$ in the expansion of $H(t_1,t_2;y_1,y_2)$ is a finite polynomial 
\bdm
B_{p,q}(y_1,y_2)=\sum_{m,n}\mu_{p,q}(m,n) y_1^m\, y_2^n \ 
\edm
giving the multiplicities of all dominant weights entering in the representation $R_{p\lambda_1+q\lambda_2}$, and thus gathering some of the most relevant information to construct the representation.

All that has been said shows that $H(t_1,t_2;y_1,y_2)$ is a generating function well worth to be computed and, as we will see, the computation can be done in a rather systematic way. The first thing to do is to use (\ref{eq:car}) to change variables in (\ref{eq:h}) in order to express $z_1$ and $z_2$ in terms of the phases $x_l$; the weights and multiplicities entering in the fundamental representations can be obtained from references such as \cite{ov90}-\cite{lie}. With this, the integrand in (\ref{eq:nmp}) takes the form of a rational function in the phases $x_l$ which can be, in general, quite involved. Notwithstanding this, the integral is very well suited to be computed by means of the residue theorem. This is so because the construction of $G(t_1,t_2; z_1,z_2)$, see \cite{nfp14}, ensures that its denominator factorizes as a product of factors with the structure $x_l^k-a$, where $k\geq 0$ and $a$ is either proportional to $t_l$ or $t_l^{-1}$. Now, we are interested in the case of small parameters, $t_s,y_s\rightarrow 0$, in order to ensure a good convergence of the series in (\ref{eq:gcar}) and, consequently, $|a|\leq 1$ only in the first case. Hence, the poles of the integrand of (\ref{eq:h}) which lie in the interior of the unit circles, and give thus contributions to the integral, can be easily identified by a simple inspection. The evaluation of the corresponding residues is a conventional task which is incorporated as a routine in most symbolic calculus languages.  So, if we choose, for instance, to integrate $x_2$ first, it is not hard to obtain 
\beq
I_2(x_1;t_1,t_2;y_1,y_2)=\sum_{p_2\in\Pi_2} {\rm Res}\left.\frac{G(t_1,t_2;z_1,z_2)}{(x_1-y_1)(x_2-y_2)}\right|_{x_2\rightarrow p_2}, \label{eq:idos}
\eeq
where $\Pi_2$ is  the set of poles with $|x_2|\leq 1$. Now, as the generating function has to be regular at the origin of the parameter space, the structure of the denominator of $I_2(x_1;t_1,t_2;y_1,y_2)$ is again of the same type, the only difference being that each $a$ incorporates now positive or negative powers of $t_l$ and $y_s$, but only of one of both signs. Thus, the identification of the poles lying inside the $x_1$ unit circle is direct and, finally
\beq
H(t_1,t_2;y_1,y_2)=\sum_{p_1 \in \Pi_1} {\rm Res}\left.I_2(x_1;t_1,t_2;y_1,y_2)\right|_{x_1\rightarrow p_1},\label{eq:iuno}
\eeq
whith the obvious definition for $\Pi_1$.

Once $H(t_1,t_2;y_1,y_2)$ has been computed, it is possible to use it to obtain  explicit formulas for the  multiplicities $\mu_{p,q}(m,n)$ of a particular dominant weight $m \lambda_1+n\lambda_2$ in the representation $R_{p\l_1+q\l_2}$  of ${\cal A}$. The idea is to extract the rational generating function $A_{m,n}(t_1,t_2)$ from $H(t_1,t_2;y_1,y_2)$ (\ref{HA}) and to decompose it in partial fractions such that the numerators are simple enough to express them as a product of two known Taylor series in $t_1$ and $t_2$. In fact, as we will see in the examples, all the series arising through this process are of the form
\beq
\frac{x^r}{(1-x^k)^s}=\sum_{p=0}^\infty \vartheta_k(p-r)\, f_s\left(\frac{p-r}{k}\right) x^p\hspace{2cm} \label{eq:series}
\eeq
for different values of $r,k$ and $s$, where
\bdm
\vartheta_k(p)=\left\{\begin{array}{lc} 1&{\rm if}\ p\geq 0\ {\rm and}\ p\equiv 0\, ({\rm mod}\; k)\\0&{\rm otherwise}\end{array}\right.\hspace{1.5cm}{\rm and}\hspace{1.5cm}f_s(q)=\left(\begin{array}{c}q+s-1\\q\end{array}\right)\ ,
\edm
and the obtention of the multiplicity formulas, though quite laborious, is feasible in this way.
\section{Results for the algebras $A_2$ and $C_2$}
We offer in this section the results for the algebras $A_2$ and $C_2\simeq B_2$. The starting points are in both cases the generating functions for characters such as they were obtained in \cite{nfp14}, see also \cite{nfp15}. For each algebra, we present the generating function $H(t_1,t_2;y_1,y_2)$ and the formulas for the multiplicities of the weights $m\lambda_1+n\lambda_2$ up to $m+n=4$. Although the calculations cannot be reproduced here in full detail, some relevant intermediate results, such as the poles which contribute to the generating function or the structure of the decomposition of the functions $A_{m,n}(t_1,t_2)$ in partial fractions, are mentioned.

A similar treatment of $G_2$ requires, as a previous step, to compute the generating function of the characters of that algebra. We defer this computation and the presentation of the results for $G_2$ until Section 4.
\subsection{Algebra $A_2$}
The generating function for characters of $A_2$ is \cite{nfp14,nfp15}
\bdm
G(t_1,t_2;z_1,z_2)=\frac{1-t_1 t_2}{(1-t_1 z_1+t_1^2 z_2-t_1^3)(1-t_2 z_2+t_2^2 z_1-t_2^3)}
\edm
and the decomposition of the fundamental characters is
\bdm
z_1=x_1+\frac1{x_2}+\frac{x_2}{x_1}\,, \quad z_2=x_2+\frac1{x_1}+\frac{x_1}{x_2}\,.
\edm
Substituting these expressions in the generating function, one finds that the set of poles of the integrand of (\ref{eq:h}) with $x_2$ inside the unit circle is $\Pi_2=\{t_1,t_2 x_1, y_2 \}$. Then, after computing $I_2(x_1;t_1,t_2;y_1,y_2)$ as in (\ref{eq:idos}), the set of poles of $I_2$ with $|x_1|\leq 1$ is identified, giving $\Pi_1=\{t_2,t_1^2,y_1,t_1 y_2\}$. Finally, the generating function is obtained through (\ref{eq:iuno}), and the result is
\bdm
H(t_1,t_2;y_1,y_2)=\frac{a_{0,0}+a_{1,0} y_1+a_{0,1} y_2+a_{1,1} y_1 y_2}
{(1 - t_1^3) (1 - t_2^3) (1 - t_1 t_2) (1 - t_1 y_1) (1 - t_2^2 y_1) 
  (1 - t_1^2 y_2) (1 - t_2 y_2)}
\edm
where
\beqrn
a_{0,0}&=&1 + t_1 t_2 + t_1^2 t_2^2,\hspace{2.3cm}
a_{1,0}=-t_1 t_2^2 (t_1^2 + t_2 + t_1 t_2^2),\\
a_{0,1}&=&-t_1^2 t_2 (t_1 + t_1^2 t_2 + t_2^2),\hspace{1.2cm}
a_{1,1}=-t_1^2 t_2^2 (1 - t_1^3 - t_1 t_2 - t_1^2 t_2^2 - t_2^3).
\eeqrn
All the generating functions $A_{m,n}(p,q)$ for $m+n\leq 4$ can be decomposed in partial fractions which can be chosen in the form
\bdm
A_{m,n}(p,q)=\frac{\widetilde{b}_{m,n}(t_1,t_2)}{(1-t_1^3)(1-t_2^3)^2}+\frac{\widetilde{c}_{m,n}(t_1,t_2)}{(1-t_1 t_2)(1-t_2^3)^2}
\edm
with $\widetilde{b}_{m,n}$ and $\widetilde{c}_{m,n}$ some polynomials with integer coefficients. Expanding these fractions using (\ref{eq:series}) and matching coefficients, we find that the weight multiplicities $\mu_{p,q}$ (\ref{Amu}) are, for $m+n\leq4$,
\beqrn
\mu_{p,q}(0,0)&=& F_{p,q}(0)-(1 + q) \delta_{p,q}\\
\mu_{p,q}(1,0)&=& F_{p,q}(2)-(1 + q) \delta_{p+2 , q} \\
\mu_{p,q}(2,0)&=&F_{p,q}(1)-(2 + q) \delta_{p+1 , q}  \\
\mu_{p,q}(1,1)&=& F_{p,q}(0)-(2 + q) \delta_{p,q}\\
\mu_{p,q}(3,0)&=& F_{p,q}(0)-(3 + q) \delta_{p,q} +  \delta_{p,0} \delta_{q,0} - 
   \delta_{p+3, q}\\
\mu_{p,q}(2,1)&=& F_{p,q}(2)-(2 + q) \delta_{p+2 , q}  - 
   \delta_{p,q+1}\\
\mu_{p,q}(4,0)&=& F_{p,q}(2)  
   -(3 + q) \delta_{p+2 , q} +\delta_{p , 1} \delta_{q,0} + \delta_{p,0} \delta_{q , 2} - 2 \delta_{p  , q+1} - \delta_{p+5 , q}\\
\mu_{p,q}(3,1)&=&F_{p,q}(1)-(3 + q) \delta_{p+1 , q}  + 
   \delta_{p,0} \delta_{q , 1} -\delta_{p , q+2} - \delta_{p+4 , q} \\
\mu_{p,q}(2,2)&=&F_{p,q}(0) -(3 + q) \delta_{p,q} + \delta_{p,0} \delta_{q,0} - 
   \delta_{p+3,q} - \delta_{p,q+3}
\eeqrn
where 
\bdm
F_{p,q}(s)=(1 + q)\, \vartheta_3(p - q+s) + (1 + p)\, \vartheta_3(q - p-s).
\edm
As a consequence of the symmetry of the Dynkin diagram of the symply-laced algebra $A_2$, the cases which do not appear explicitly  in the list can be obtained through the formula $\mu_{p,q}(m,n)=\mu_{q,p}(n,m)$.

Notice also that in each of the  formulas for the weight multiplicities a summand  of the form $F_{p,q}(s)-(1+q)\delta_{p+s,q}$, for $s=0$, 1 or 2, appears; as it is easy to check, we have that $F_{p,q}(s)-(1+q)\delta_{p+s,q}=(\min(p,q)+1)\,\widehat\vartheta_3(p-q+s)$, where $\widehat\vartheta_3(u)=\vartheta_3(|u|)$, $\forall u\in\bf Z$, and the formulas above can be expressed equally in terms of $\min(p,q)$. For instance, $\mu_{p,q}(1,1)=(\min(p,q)+1)\,\widehat\vartheta_3(p-q)-\delta_{p,q}$, $\mu_{p,q}(2,0)=(\min(p,q)+1)\,\widehat\vartheta_3(p-q+1)-\delta_{p+1,q}$, $\mu_{p,q}(1,0)=(\min(p,q)+1)\,\widehat\vartheta_3(p-q+2)-\delta_{p+2,q}$, and so on.

\subsection{Algebra $C_2$}
The generating function for characters is in this case \cite{nfp14,nfp15}
\bdm
G(t_1,t_2;z_1,z_2)=\frac{1+t_2-z_1 t_1 t_2+t_1^2 t_2+t_1^2 t_2^2}{(1-(t_1+t_1^3) z_1+t_1^2 (z_2+1)+t_1^4)(1-(t_2+t_2^3)(z_2-1)+t_2^2(z_1^2-2z_2)+t_2^4)} 
\edm
whith the fundamental characters given by
\bdm
z_1=x_1 +\frac{1}{x_1}+\frac{x_1}{x_2}+\frac{x_2}{x_1}\,,\quad z_2
=1+x_2+\frac{1}{x_2}+\frac{x_1^2}{x_2}+\frac{x_2}{x_1^2} \,.
\edm
The two sets of poles contributing to the generating function $H$ are 
\bdm
\Pi_2=\{t_2,t_1 x_1,t_2 x_1^2, y_2\},\hspace{1cm}\Pi_1=\{\pm\sqrt{t_2 y_2},\pm t_2,t_1,t_1 y_2,y_1\}
\edm
and the result turns out to be
\bdm
H(t_1,t_2;y_1,y_2)=\frac{c_{0,0}+c_{1,0}\, y_1+c_{0,1}\, y_2+c_{1,1}\,y_1 y_2+c_{2,0} \,y_1^2 +c_{2,1} \,y_1^2 y_2}
{(1 - t_1^2)^2(1 - t_2^2)(1 - t_2)(1 - t_1y_1)(1 - t_2^2y_1^2)(1 - t_1^2y_2)(1 - t_2y_2)},
\edm
where
\beqrn
c_{0,0}&=&1+t_1^2 t_2,\hspace{1.75cm}c_{1,0}=t_1 t_2(1-t_1^2),\hspace{1.3cm} c_{0,1}=-t_1 t_2(t_1^3+t_1 t_2),\\
c_{1,1}&=&t_1 t_2 (t_1^4-t_1^2),\hspace{1cm} c_{2,0}=-t_1^2 t_2^2(1+t_2),\hspace{1cm}c_{2,1}
=t_1^2 t_2^2(t_1^2+t_1^2 t_2+t_2^2-1).
\eeqrn
The generating functions $A_{m,n}(p,q)$ can be decomposed in partial fractions according to the form
\bdm
A_{m,n}(p,q)=\frac{\widetilde{f}_{m,n}(t_1,t_2)}{2(1-t_1^2)^2(1-t_2)^2}+\frac{\widetilde{g}_{m,n}(t_1,t_2)}{2(1-t_1^2)(1-t_2^2)}
\edm
and the resulting multiplicity formulas are,   when $m+n\leq4$,
\beqrn
\mu_{p,q}(0,0)&=&\frac{1}{2} \vartheta_2(p) (M + \vartheta_2(q))\\
[4pt]
\mu_{p,q}(1,0)&=&\frac{1}{2}  \vartheta_2(p-1) M\\
[4pt]
\mu_{p,q}(0,1)&=&\frac{1}{2} \vartheta_2(p) (M - \vartheta_2(q))\\
[4pt]
\mu_{p,q}(2,0)&=&\frac{1}{2} \vartheta_2(p) (M - 2 + \vartheta_2(q))\\
[4pt]
\mu_{p,q}(1,1)&=&\frac{1}{2} \vartheta_2(p-1) (M - 2)\\
[4pt]
\mu_{p,q}(0,2)&=&\frac{1}{2} \vartheta_2(p) (M - 4 + \vartheta_2(q)) + \delta_{p,0}\\
[4pt]
\mu_{p,q}(3,0)&=&\frac{1}{2} \vartheta_2(p-1) (M - 4 + 2 \delta_{q,0})\\
[4pt]
\mu_{p,q}(2,1)&=&\frac{1}{2} \vartheta_2(p) (M - 4 + 2 \delta_{q,0} - \vartheta_2(q)) + \delta_{p,0}\\
[4pt]
\mu_{p,q}(1,2)&=&\frac{1}{2} \vartheta_2(p-1) (M - 6 + 2 \delta_{q,0}) + \delta_{p,1}\\
[4pt]
\mu_{p,q}(0,3)&=&
  \frac{1}{2} \vartheta_2(p) (M - 8 + 4 \delta_{q,0} - \vartheta_2(q)) + 3 \delta_{p,0} + \delta_{p,2} - \delta_{p,0} \delta_{q,0}\\
  [4pt]
\mu_{p,q}(4,0)&=&
  \frac{1}{2} \vartheta_2(p) (M - 8 + 2 \delta_{q,1} + 4 \delta_{q,0} + \vartheta_2(q)) + 2 \delta_{p,0} - \delta_{p,0} \delta_{q,0}\\
  [4pt]
\mu_{p,q}(3,1)&=&\frac{1}{2} \vartheta_2(p-1) (M - 8 + 2 \delta_{q,1} + 4 \delta_{q,0}) + \delta_{p,1}\\
[4pt]
\mu_{p,q}(2,2)&=&
  \frac{1}{2} \vartheta_2(p) (M - 10 + 2 \delta_{q,1} + 4 \delta_{q,0} + \vartheta_2(q)) + 3 \delta_{p,0} + \delta_{p,2} - \delta_{p,0} \delta_{q,0}\\
  [4pt]
\mu_{p,q}(1,3)&=&
  \frac{1}{2} \vartheta_2(p-1)(M - 12 + 2 \delta_{q,1} + 6 \delta_{q,0}) + 3 \delta_{p,1} + \delta_{p,3} - 
   \delta_{p,1} \delta_{q,0}\\
   [4pt]
\mu_{p,q}(0,4)&=&
  \frac{1}{2} \vartheta_2(p) (M - 16 + 4 \delta_{1,q} + 8 \delta_{q,0} + \vartheta_2(q)) + 6 \delta_{p,0} + 3 \delta_{p,2} + 
   \delta_{p,4} - 3 \delta_{p,0} \delta_{q,0} \\&\ & -\ \delta_{p,0} \delta_{q,1} - \delta_{p,2} \delta_{q,0}
\eeqrn
where we have written $M=(p+1)(q+1)$.
\section{Results for the algebra $G_2$}
\subsection{The generating function of the characters of $G_2$}
First of all, we shall compute the generating function of the characters of $G_2$ following the four-step method described in \cite{nfp14}. The relevant information on the root and weight systems and the Clebsch-Gordan series of the algebra is taken from standard references such as \cite{ov90}-\cite{lie}. The background on the quantum Calogero-Sutherland model needed for the calculation is explained in \cite{op83} or in several other references cited in \cite{nfp14}. 

The exceptional algebra $G_2$ has positive roots
\bdm
\alpha_1,\ \ \alpha_2,\ \  \alpha_1+\alpha_2,\ \  2\alpha_1+\alpha_2,\ \ 3\alpha_1+\alpha_2,\ \ 3\alpha_1+ 2\alpha_2 
\edm
and Cartan matrix
\bdm
A=\left(\begin{array}{cc}2&-3\\-1&2\end{array}\right).
\edm
The two fundamental weights are
\bdm
\lambda_1=2\alpha_1+\alpha_2,\ \ \ \ \lambda_2=3\alpha_1+ 2\alpha_2
\edm
and the dimension of the representation $R_{p \lambda_1+q \lambda_2}$ is given by the Weyl formula
\bdm
\dim(p, q) = 
    \frac{1}{120}(p+1)(q+1)(p + q+2)(p + 2 q+3)(p + 3q+4)(2p + 3q+5) .
\edm
The first thing we have to do is to write the quantum Calogero-Sutherland Hamiltonian $\Delta_z^1$, i.e with coupling constants equal to one, using as coordinates the fundamental characters $z_1$ and $z_2$. In order to do so, we recall that the eigenfunctions of the Hamiltonian are the characters of $G_2$, and the eigenvalues are
\bdm
\varepsilon(p, q;1)= 4 p^2 + 12 q^2 + 12 p\, q+20 p + 36 q .
\edm
On the other hand, the $G_2$ Clebsch-Gordan series
\beqrn
z_1^2&=&\bchi_{2, 0} + z_1 + z_2 + 1\\
z_1 z_2&=&\bchi_{1, 1} + \bchi_{2, 0} + z_1\\
z_2^2&=&\bchi_{0, 2} + \bchi_{3, 0} + \bchi_{2, 0} + z_2 + 1\\
z_1 \bchi_{2, 0}&=&\bchi_{3, 0} + \bchi_{1, 1} + \bchi_{2, 0} + z_1 + z_2
\eeqrn
allow us to solve for the quadratic characters
\beqrn
\bchi_{2, 0}&=& z_1^2 - z_2-z_1-1\\
\bchi_{1, 1}&=&z_1 z_2 - z_1^2 + z_2 +1\\
\bchi_{0, 2}&=& z_2^2- z_1^3+2 z_1 z_2 + z_2 + 2 z_1
\eeqrn
and using all these data in the Schr\"{o}dinger equation the form of the Hamiltonian is univocally fixed to be
\beqr
\Delta_z^1&=&4(z_1^2-z_2-4z_1-7)\,\partial_{z_1}^2+4(3z_2^2-3z_1^3+6z_1 z_2-5z_1^2+2z_2+11z_1-7)\,\partial_{z_2}^2\nonumber\\&&+\ 4(3z_1 z_2-7z_1^2+7z_2-8z_1+7)\,\partial_{z_1} \partial_{z_2}+24 z_1 \partial_{z_1}+48 z_2 \partial_{z_2}\, .\label{eq:ham}
\eeqr
The differential equation for the generating function for characters $G(t_1,t_2;z_1,z_2)$ is thus
\beq
(\Delta_t-\Delta_z)G(t_1,t_2;z_1,z_2)=0, \label{eq:dif}
\eeq
where $\Delta_t$ comes from the Hamiltonian eigenvalues and its explicit form is
\bdm
\Delta_t=4\; t_1^2\; \partial_{t_1}^2+12\; t_2^2\; \partial_{t_2}^2+12\; t_1 t_2\; 
    \partial_{t_1}\partial_{t_2}+24\; t_1\; \partial_{t_1}+48\; t_2\; \partial_{t_2} .
\edm
With this setup, we proceed to solve (\ref{eq:dif}) through the following four steps:\\

\noindent{\bf Step (i):} The dimensions of $R_{\lambda_1}$ and $R_{\lambda_2}$ are 7 and 14, and the weights entering in these representations are, respectively
\beqrn
  &&\{0,\ \pm\lambda_1,\ \pm(2\lambda_1-\lambda_2),\ \pm(\lambda_1-\lambda_2)\}\\
  &&\{0,\ 0,\ \pm\lambda_2,\  \pm\lambda_1,\  \pm(3\lambda_1-\lambda_2),\ \pm(\lambda_1-\lambda_2),\ \pm(3\lambda_1-2\lambda_2),\ \pm(2\lambda_1-\lambda_2)\}. 
\eeqrn
This gives the fundamental characters as
\beqr
z_1 &=& 1 + x_1 + \frac{1} {x_1} + \frac{x_1}{x_2} + \frac{x_2}{x_1} + \frac{x_1^2}{x_2} + \frac{x_2}{x_1^2} \label{eq:xg21}\\ 
z_2 &=& 2 + x_1 + \frac{1}{x_1} + x_2 + \frac{1}{x_2} + \frac{x_1}{x_2} + \frac{x_2}{x_1} + \frac{x_1^2}{x_2} + \frac{x_2}{x_1^2}+ 
    \frac{x_1^3}{x_2} +\frac{x_2}{x_1^3}+ \frac{x_1^3}{x_2^2} + \frac{x_2^2}{x_1^3}\label{eq:xg22}.
\eeqr
The Weyl orbits of the fundamental weights are, on the other hand
\beqrn
W\cdot \lambda_1&=&\{\pm\lambda_1,\ \pm(2\lambda_1-\lambda_2),\ \pm(\lambda_1-\lambda_2)\}\\
W\cdot \lambda_2&=&\{\pm\lambda_2,\ \pm(3\lambda_1-\lambda_2),\ \pm(3\lambda_1-2\lambda_2) \}
\eeqrn
and the denominator of the generating function for characters is thus
\bdm
D(t_1,t_2;z_1,z_2)=D_1\times D_2,
\edm
with
\beqrn
D_1&=&1 + t_1^6 + (t_1+t_1^5) (1 - z_1)  + (t_1^2+t_1^4)(1 + z_2)  + 
  t_1^3 (1 - z_1^2 + 2 z_2) \\
  [4pt]
D_2&=&1 + t_2^6 + (t_2+t_2^5)(1 + z_1 - z_2) + 
  (t_2^2+t_2^4) (1 - z_1 + z_1^3 - 2 z_2 - 3 z_1 z_2)\\ &&
 +\  t_2^3 (1 - 2 z_1 - z_1^2 + 2 z_1^3 - 4 z_2 - 4 z_1 z_2 - z_2^2).
\eeqrn

\noindent{\bf Step (ii):} Taking $z_1=7$ and $z_2=14$ in the above denominator and using the Weyl formula for dimensions, it is not difficult to obtain the generating function $E(t_1,t_2)$ of the dimensions of the representations of the algebra $G_2$. It turns out to be
\bdm
E(t_1,t_2)=\frac{P(t_1,t_2)}{(1-t_1)^6(1-t_2)^6}
\edm
where
\beqrn
P(t_1,t_2)&=&1 + t_1 + 8 t_2 - 26 t_1 t_2 + 
    15 t_1^2 t_2 - 6 t_1^3 t_2 + 
    t_1^4 t_2 + 8 t_2^2 - 41 t_1 t_2^2 + 
    78 t_1^2 t_2^2 \\&& -\  41 t_1^3 t_2^2 + 
    8 t_1^4 t_2^2 + t_2^3 - 6 t_1 t_2^3 + 
    15 t_1^2 t_2^3 - 26 t_1^3 t_2^3 + 
    8 t_1^4 t_2^3 + t_1^3 t_2^4 + 
    t_1^4 t_2^4 .
\eeqrn

\noindent{\bf Step (iii):} The characters of $G_2$ can be computed by solving the Schr\"{o}dinger equation for the Calogero-Sutherland Hamiltonian (\ref{eq:ham}). By means of a few of these characters, and taking into account the monomials appearing in $P(t_1,t_2)$, one can conjecture that the numerator $N(t_1,t_2;z_1,z_2)$ of the generating function of characters is
\beqrn
N(t_1,t_2;z_1,z_2)&=&1 + t_1 + t_1^4 t_2 + t_2^3 + t_1^3 t_2^4 + t_1^4 t_2^4 +(1 - z_1)(t_1^3 t_2+t_1t_2^3)\\ 
&&+\ (1 + z_1)(t_2+t_2^2+t_1^4 t_2^2+t_1^4 t_2^3)+(1 + z_1 - z_1^2)(t_1 t_2^2+t_1^3 t_2^2)\\
&&
+\  (1+z_2)(t_1^2 t_2+t_1^2 t_2^3) +
(2 + z_1 - z_1^2 + z_2)(t_1 t_2+t_1^3 t_2^3)   \\
&&
+\ (1 + 2 z_1 - z_1^2 + z_2 + z_1 z_2) t_1^2 t_2^2.
\eeqrn
\noindent{\bf Step (iv):} It remains to verify that the conjecture is correct and
\beq
G(t_1,t_2;z_1,z_2)=\frac{N(t_1,t_2;z_1,z_2)}{D(t_1,t_2;z_1,z_2)} \label{eq:gg2}
\eeq
does indeed satisfy the differential equation (\ref{eq:dif}). This is in fact the case, as one can check by direct substitution in (\ref{eq:dif}). Thus (\ref{eq:gg2}) correctly gives the generating function for characters of the algebra $G_2$ 
\subsection{The generating function $H$ and multiplicity formulas for $G_2$}
Substitution of the decompositions (\ref{eq:xg21}), (\ref{eq:xg22}) in the generating function (\ref{eq:gg2}), allows to identify the relevant poles for computing the function $H$. They are 
\beqrn
\Pi_2&=&\{t_1 x_1,t_1 x_1^2,t_2,t_2 x_1^3, \pm\sqrt{t_2 x_1^3},y_2\}\\
[3pt]
\Pi_1&=&\{t_1, t_1^2,t_2, \sqrt[3]{t_2^2},\sqrt[3]{t_2^2}\, e^{\pm i \frac{2\pi}{3}}, t_1 y_2, \pm\sqrt{t_1 y_2},\sqrt[3]{t_2 y_2}, \sqrt[3]{t_2 y_2}\, e^{\pm i \frac{2\pi}{3}},\sqrt[3]{t_2 y_2^2}, \sqrt[3]{t_2 y_2^2}\, e^{\pm i \frac{2\pi}{3}},y_1 \}. 
\eeqrn
After the residues are computed, one finds that the $H$ function has the form
\beq
\label{HG2}
H=\frac1D{\sum_{r=0}^5\sum_{s=0}^3 g_{r,s} \,y_1^r y_2^s}
\eeq
where the denominator is
\beqrn
D&=&(1 - t_1)^2 (1 - t_1^2) (1 - t_1^3) (1 - t_2)^3 (1 - t_2^2) (1 - t_1 y_1) 
  (1 - t_1^2 y_1)  \\&&\times\ (1 - t_2 y_1)(1 - t_2^2 y_1^3) (1 - t_1^2 y_2) (1 - t_1^3 y_2) 
  (1 - t_2 y_2) (1 - t_2^2 y_2)
\eeqrn
and the 24 coefficients $g_{r,s}$ in the numerator, which are too long to be written here, can be read from the adjoint file {\tt NumeratorHG2.txt}, where they were written in text format in order that the file can be easily used in any program for symbolic computations.  
The decomposition of $A_{m,n}(t_1,t_2)$ in partial fractions has the structure
\bdm
A_{m,n}(p,q)=\frac{\widetilde{h}_{m,n}(t_1,t_2)}{9(1-t_1^3)(1-t_2)^2}+\frac{\widetilde{j}_{m,n}(t_1,t_2)}{8(1-t_1^2)(1-t_2^2)}+\frac{\widetilde{k}_{m,n}(t_1,t_2)}{72(1-t_1)^4(1-t_2)^4}
\edm
and the multiplicities are, when $m+n\leq4$,
\beqrn
\mu_{p,q}(0,0)&=& F_{p,q}(29,2,3)\\
[4pt]
\mu_{p,q}(1,0)&=& F_{p,q}(17,-1,-1)\\
[4pt]
\mu_{p,q}(0,1)&=& F_{p,q}(-7,2,-1)\\
[4pt]
\mu_{p,q}(2,0)&=& F_{p,q}(-19,-1,3)\\
[4pt]
\mu_{p,q}(1,1)&=& F_{p,q}(-55,-1,-1)+1\\
[4pt]
\mu_{p,q}(0,2)&=& F_{p,q}(-115,2,3)+4-2\delta_{p,0}-\delta_{q,0}\\
[4pt]
\mu_{p,q}(3,0)&=& F_{p,q}(-79,2,-1)+2 - \delta_{p,0}\\
[4pt]
\mu_{p,q}(2,1)&=& F_{p,q}(-127,-1,-1)+5 - \delta_{p,1} - 2 \delta_{p,0} - \delta_{q,0}\\
[4pt]
\mu_{p,q}(1,2)&=& F_{p,q}(-199,-1,-1)+12 - \delta_{p,2} - 3 \delta_{p,1} - 6 \delta_{p,0} - \delta_{q,1} - 4 \delta_{q,0} + \delta_{p,0} \delta_{q,0}\\
[4pt]
\mu_{p,q}(0,3)&=& F_{p,q}(-295,2,-1)+26 - 2 \delta_{p,3} - 4 \delta_{p,2} - 8 \delta_{p,1} - 16 \delta_{p,0} -\delta_{q,2} - 4 \delta_{q,1} + 
  \delta_{p,0} \delta_{q,1}\\&\ &+\ (-11 + \delta_{p,1} + 5 \delta_{p,0}) \delta_{q,0}\\
  [4pt]
\mu_{p,q}(4,0)&=& F_{p,q}(-163,-1,3)+8- \delta_{p,2} - 
  2 (\delta_{p,1} + 2 \delta_{p,0} + \delta_{q,0})\\
  [4pt]
\mu_{p,q}(3,1)&=& F_{p,q}(-223,2,-1)+15 - \delta_{p,3} - 2 \delta_{p,2} - 4 \delta_{p,1} - 9 \delta_{p,0} - \delta_{q,1} - 5 \delta_{q,0} + 2 \delta_{p,0} \delta_{q,0}\\
[4pt]
\mu_{p,q}(2,2)&=& F_{p,q}(-307,-1,3)+28 - \delta_{p,4} - 2 \delta_{p,3} - 5 \delta_{p,2} - 10 \delta_{p,1} - 17 \delta_{p,0} - \delta_{q,2}  + 
  (\delta_{p,0}-4) \delta_{q,1}\\ && +\ (2 \delta_{p,1}  + 5 \delta_{p,0}-12) \delta_{q,0}\\
  [4pt]
\mu_{p,q}(1,3)&=& F_{p,q}(-415,-1,-1)+51 - \delta_{p,5} - 3 \delta_{p,4} - 6 \delta_{p,3} - 12 \delta_{p,2} - 21 \delta_{p,1} - \delta_{q,3} + 
  \delta_{p,0} (\delta_{q,2}-33)\\ &\ & -\ 4 \delta_{q,2}  + (\delta_{p,1} + 4 \delta_{p,0}-11) \delta_{q,1}  + (2 \delta_{p,2} + 6 \delta_{p,1} + 13 \delta_{p,0}-25) \delta_{q,0}\\
  [4pt]
\mu_{p,q}(0,4)&=& F_{p,q}(-547,2,3)+88 - 2 \delta_{p,6} - 4 \delta_{p,5} - 8 \delta_{p,4} - 16 \delta_{p,3} - 26 \delta_{p,2} - 40 \delta_{p,1} - 
  \delta_{q,4}-4 \delta_{q,3} \\&\ &  +\  \delta_{p,0} (\delta_{q,3}-62) + 
  (\delta_{p,1} + 4 \delta_{p,0}-11) \delta_{q,2}  + 
  (\delta_{p,2} + 4 \delta_{p,1} + 12 \delta_{p,0}-24) \delta_{q,1} \\ &&
 +\  (3 \delta_{p,3} + 7 \delta_{p,2} + 15 \delta_{p,1} + 29 \delta_{p,0}-48) \delta_{q,0}
\eeqrn
where the functions $F_{p,q}(r,s,t)$ are given as
\beqrn
F_{p,q}(r,s,t)&=& \frac{1}{72} (p+1)(q+1)\left(r+2 p^2+6 q^2+9 p\,q+13 p+21 q\right)\\&&+\ \frac{s}{9} (q+1)\left(\vartheta_3(p)-\vartheta_3(p-1)\right)+\frac{t}{8}\vartheta_2(p)\,\vartheta_2(q) 
\eeqrn
and can be obtained from a generating function $\widetilde F(r,s,t;t_1,t_2) = \sum_{p,q=0}^\infty F_{p,q}(r,s,t)t_1^pt_2^q$ of the form
\beqrn
\widetilde F(r,s,t;t_1,t_2)& =&  \frac{r}{72(1 - t_1)^2 (1 - t_2)^2} + \frac{(9 t_2 - 3) t_1^2  + (5 t_2^2 - 22 t_2 + 5) t_1  - 3  (t_2 - 3) t_2}{12 (1 - t_1)^4 (1 - t_2)^4}\\[4pt]&&+\ 
  \frac{ 8 s (1 - t_1^2) (1 + t_2) + 9 t (1 + t_1 + t_1^2) (1 - t_2)}{72 (1 - t_1^2) (1 + t_1 + t_1^2) (1 - t_2)^2 (1 + t_2)} .
\eeqrn

We have verified that  our formulas give accurate values for the weight multiplicities in the rank of  $p$ and $q$ where there are available sources to compare, see \cite{lie}.
On the other hand, as far as we know, the only other reference where the problem of finding the generating function for the weight multiplicities for $G_2$ is addressed is a paper by Dokovi\'c \cite{dok}. However, upon comparison with our results and \cite{lie}, it seems that the generating functions in that paper, maybe due to misprints, are not entirely correct and yield the wrong results for some  multiplicities.
\section*{Acknowledgements}
J.F.N. and W.G.F. acknowledge financial support from, respectively, MTM2012-33575 and MTM2014-57129 projects, SGPI-DGICT(MEC), Spain.

\end{document}